\documentstyle[aps,eqsecnum,prc]{revtex}
\begin{document}
\title{
Quasiscaling in the analysis of the yield ratio $\pi^-/\pi^+$:\\
Mathematical structure and estimation of source size}
\author{Takeshi Osada,
\thanks{e-mail: osada@azusa.shinshu-u.ac.jp.} 
Minoru Biyajima 
\thanks{e-mail: minoru44@jpnyitp.bitnet}
and Grzegorz Wilk 
\thanks{Permanent address:
Soltan Institute for Nuclear Studies, Zd-PVIII, Ho\.za 69, 
PL-00-681 Warsaw, Poland}
\thanks{e-mail: wilk@fuw.edu.pl.}}
\address{Department of Physics, Faculty of Science, 
         Shinshu University, Matsumoto 390, Japan}

\date{\today}

\maketitle
\begin{abstract}
Recently we have found 
that integral of the squared Coulomb wave function describing system
composed of charged pion and central charged fragment 
$Z_{\mbox{\scriptsize eff}}$ protons, $|\psi_r(r)|^2$, times pion
source function $\rho(r)$ (of the  size $\beta$), $\int
dr~|\psi_r(r)|^2 \rho(r)$, shows a quasiscaling behavior. 
This is approximately invariant under the following transformation:
$(\beta,Z_{\mbox{\scriptsize eff}}) \to (\lambda\beta,\lambda
Z_{\mbox{\scriptsize eff}})$; $\lambda >0$. We called such behavior
$\beta$-$Z_{\mbox{\scriptsize eff}}$ quasiscaling. 
We examine this quasiscaling behavior in detail. In 
particular we provide a semi-analytical examination of this behavior and
confirm it for the exponential pionic source functions in addition to the
Gaussian ones and for the production of $K$ mesons as well. When
combined with the results of the HBT, a result of the yield ratio 
allows us to estimate the size of the central charged fragment (CCF) 
to be $125\le Z_{\mbox{\scriptsize eff}}\le 150$ for Pb+Pb collisions
at energy $158$ GeV/nucleon.  From our estimation, the baryon number 
density $0.024 \le n_{\mbox{\scriptsize B}}\le0.036$ [1/fm$^3$] 
is obtained.
\end{abstract}

\section{Introduction}
Recently E866 \cite{E866AuAu} and NA44 \cite{NA44PbPb} collaborations
reported a remarkable deviation from unity of the yield ratio
$\pi^-/\pi^+$ measured at small $m_{\mbox{\scriptsize T}}$-$m_{\pi}$
region in high energy heavy-ion collisions. Gyulassy and Kauffmann
\cite{GYUL-KAUFF} have argued that such deviations 
are due to the Coulomb 
interaction between charged $\pi^{\pm}$ and a system of 
$Z_{\mbox{\scriptsize eff}}$ protons (called a central charged
fragment - CCF) which are nearly stopped 
in the central rapidity region 
in the course of the heavy ion collision. They have provided 
theoretical formulae for this effect calculated up to 
the first order in the $Z_{\mbox{\scriptsize eff}}\alpha$. 
On the other hand Shuryak\cite{Shuryak} has 
also pointed out that the above deviation  seen in
the high energy nucleus-nucleus collisions can be regarded as a
manifestation of a Coulomb interaction effect and have proposed a
theoretical formula using the Gamow factor. 
See also refs.\cite{LI,WONG}\\ 

In our previous paper \cite{YIELD-RATIO} we have derived a 
new theoretical 
formula for the observed yield ratio $\pi^-/\pi^+$ using the Coulomb
wave function. We have found that, when applied to the experimental
data, this formula exhibits a valley like structure of the minimum
$\chi^2$ values in $\beta$-$Z_{\mbox{\scriptsize eff}}$ parameter space. 
This result suggests existence of the following
$\beta$-$Z_{\mbox{\scriptsize eff}}$ quasiscaling in the pion 
production yield $N^{\pi^{\pm}}$ \cite{YIELD-RATIO}:
\begin{eqnarray}
N^{\pi^{\pm}}(m_{\scriptscriptstyle T}-m_{\pi}; 
~\lambda\!\times\!\eta_{\pm},~\lambda\!\times\!\beta)
~\approx~ N^{\pi^{\pm}}(m_{\scriptscriptstyle T}-m_{\pi}
;~\eta_{\pm},~\beta)\label{SCALING}, 
\end{eqnarray} 
where $\eta_{\pm}=\pm Z_{\mbox{\scriptsize eff}}~\mu\alpha /\mbox{p}_{r}$
and $N^{\pi^{\pm}}$ is given by \cite{FOOT1}
\begin{eqnarray}
&&N^{\pi^{\pm}}(m_t-m_{\pi};\eta_{\pm},\beta)
~\Big|_{\mbox{\scriptsize fixed $y_{\pi}$}}
=\int_{}^{}d^3{\bf r}~\rho({\bf r})~\Big|
\psi_r^{\pm}({\bf p}_r,{\bf r})\Big|^2 \nonumber \\ 
&&\quad=G(\eta_{\pm})\sum_{n=0}^{\infty}
\sum_{m=0}^{\infty}\frac{(-i)^n(i)^m}{n+m+1}
~A_n(\eta_{\pm})A_m^*(\eta_{\pm})~
I_{\beta}(n,m)~|~2{\bf p}_{r}~|^{n+m}\label{WAVE-INTEGRAL}. 
\end{eqnarray}
Here $G(\eta_{\pm})=2\pi\eta_{\pm}/\big(\exp(2\pi\eta_{\pm})-1\big)$
is the Gamow factor and 
\begin{eqnarray} 
A_n(\eta_{\pm})\!&=&\!\frac{\Gamma(i\eta_{\pm}+n)}
{\Gamma(i\eta_{\pm})~(n!)^2}~,  \quad 
I_{\beta}(n,m)=4\pi
\displaystyle{\int^{}_{}\!dr~r^{m+m+2} \rho(r)}~. \label{AI}
\end{eqnarray}
The Coulomb wave function $\psi_r^{\pm}({\bf p}_{r},{\bf r})$ of the
CCF system of $Z_{\mbox{\scriptsize eff}}$ protons and a charged
$\pi^{\pm}$ is a solution of the following Schr\"odinger equation:
\begin{eqnarray}
\bigg[~~\frac{\widehat{{\bf p}}^2_r}{2\mu}
\pm \frac{Z_{\mbox{\scriptsize eff}}~e^2}{r}~~\bigg]\psi^{\pm}_r({\bf r})
=E_{r}\psi^{\pm}_r({\bf r}),
\end{eqnarray} 
and is given explicitly in terms of the confluent hypergeometric 
function $F$~\cite{Schiff}:
\begin{eqnarray}
\psi_r^{\pm}({\bf p}_{r},{\bf r})=
\Gamma(1+i\eta_{\pm})e^{-\pi\eta_{\pm}/2}e^{i{\bf p}_r\cdot{\bf r}}
F\big(-i\eta_{\pm},1,i~(\mbox{p}_{r}r-{\bf p}_{r}\cdot{\bf r})\big)~ 
\label{CONFHYP}. 
\end{eqnarray}
For the Gaussian source, $\rho(r)=\left(\frac{1}{\sqrt{2\pi}\beta}\right)^3
\exp\left(\frac{-r^2}{ 2\beta^2}\right)$ and 
for the exponential source, $\rho(r)=\frac{1}{8\pi\beta^3}
\exp\left(\frac{-r}{\beta}\right)$, we have 
$I_{\beta}=\frac{2}{\sqrt{\pi}}(\sqrt{2}\beta)^{n+m}
~\Gamma(\frac{n+m+3}{2})$ and 
$I_{\beta}=\frac{1}{2}\beta^{n+m}\Gamma(n+m+3)$, respectively.
In the next section we demonstrate that the quasiscaling behavior is
independent of the type of the pionic source used. In Section III we
examine in detail the quasiscaling behavior of both
$N^{\pi^{\pm}}$ and $N^{K^{\pm}}$ production yields. 
In particular, we provide both semi-analytical and
numerical proofs of this behavior. In Section IV 
we estimate the most probably size $Z_{\mbox{\scriptsize eff}}$ of the CCF 
and baryon number density of it using results of tht HBT analysis. 
In the last section we present our concluding remarks.\\ 

\section{Analysis of data by NA44 Collaboration 
using exponential source function}
We analyze data of the yield ratio $\pi^-/\pi^+$ observed in Pb+Pb 
collisions at 158 GeV/nucleon\cite{NA44PbPb} by applying the 
following formula, 
\begin{eqnarray}
\pi^-/\pi^+=
\frac{N^{\pi^{-}}(m_{\scriptscriptstyle T}-m_{\pi};\eta_-,\beta)}
{N^{\pi^{+}}(m_{\scriptscriptstyle T}-m_{\pi};\eta_+,\beta)}. \label{FINAL}
\end{eqnarray}
The exponential source function in $N^{\pi^{\pm}}$ is used 
in eq.(\ref{FINAL}).
In Fig.1 we present the result by this 
source function and the previous result in ref.\cite{YIELD-RATIO}.
In both cases we fix values of $\beta$ and estimate
$Z_{\mbox{\scriptsize eff}}$ providing the minimal value of $\chi^2$
(by using CERN-MINUIT program). In both cases the corresponding
parameters $Z_{\mbox{\scriptsize eff}}$ and $\beta$ are 
proportional to each other: $Z_{\mbox{\scriptsize eff}} \approx
\kappa\beta$ (where $\kappa$ is a constant depending on the type of source
function used). In this way we confirm that the quasiscaling
behavior is independent of the type of source function used.\\

\section{Examination of the quasiscaling behavior of $N^{\pi^{\pm}}$
and $N^{K^{\pm}}$}
\subsection{Taylor series expansion and finite difference method}
We elucidate the mathematical structure of the
quasiscaling behavior seen in Fig. 1. The numerical results
presented there suggest that the series expansion part of
eq.(\ref{WAVE-INTEGRAL}) with arguments $\lambda\eta_{\pm}$ and
$\lambda\beta$ has the same $\lambda$ dependence as the $1/G(\lambda
\eta_{\pm})$ part, i.e., that 
\begin{eqnarray} 
S(\lambda~\eta_{\pm}, \lambda~\beta) \!\!&\equiv&\!\!
\sum_{n=0}^{\infty}\sum_{m=0}^{\infty}\frac{(-i)^n(i)^m}{n+m+1}
~A_n(\lambda \eta_{\pm})A_m^*(\lambda \eta_{\pm})~
I_{(\lambda\beta)}(n,m)~|~2{\bf p}_{r}~|^{n+m} \nonumber \\
\!\!&\approx&\!\! \mbox{constant}\times 1/G(\lambda \eta_{\pm}).
\label{LAMBDA-EXPAN} 
\end{eqnarray} 
\noindent
To examine this supposition we expand the
$S(\lambda\eta_{+},\lambda\beta)$ (with the Gaussian source function) and
$1/G(\lambda \eta_{+})$ in the Taylor series around $\lambda_0~(=1,2)$
with fixed $m_{\scriptscriptstyle T}$-$m_{\pi}$(=~50, 100 and 200 MeV) 
by using the {\it Mathematica} software. All combinations of
indices $(n,m)$ satisfying $0 \le n+m \le 90$ were considered. The
result is presented in Table \ref{tab1}\cite{FOOT2}. One can observe that
coefficients of $\lambda$ expansion in $1/G(\lambda \eta_{\pm})$
agree reasonably with those emerging from
$S(\lambda\eta_{+},\lambda\beta)$ (modulus constant multiplying
$S(\lambda\eta_{+},\lambda\beta)$.).\\

However, it is difficult to evaluate higher order terms of the series
in $S(\lambda\eta_+,\lambda\beta)$ using this method \cite{FOOT4}.
Therefore, we also calculate five coefficients $c_j$ ($0\le j
\le4$) of the $\lambda$-expansion in
$S(\lambda\eta_{+},\lambda\beta)$ and $1/G(\lambda \eta_{+})$ using
the finite difference method (with its size equal to
$\varepsilon=1/200$) defined as follows :  
\begin{eqnarray}
F(\lambda)&=&c_0+c_1~\delta\lambda+c_2~\delta\lambda^2
+c_3~\delta\lambda^3+c_4~\delta\lambda^4+\cdots \label{FD-METHOD},\\
&&\hspace*{-1.70cm}\mbox{where } \delta\lambda = \mid\lambda -
\lambda_0\mid~\ll~\varepsilon \mbox{~~and }\nonumber,\\
c_0&=&F(\lambda_0) \nonumber,\\
c_1&=&\frac{1}{1!\varepsilon}\big[
F(\lambda_0+\varepsilon/2)-F(\lambda_0-\varepsilon/2)\big]\nonumber, \\
c_2&=&\frac{1}{2!\varepsilon^2}\big[
F(\lambda_0+\varepsilon)-2~F(\lambda_0)+F(\lambda_0-\varepsilon)
\big]\nonumber \nonumber,\\
c_3&=&\frac{1}{3!\varepsilon^3}\big[
F(\lambda_0+3\varepsilon/2)-3~F(\lambda_0+\varepsilon/2)
+3~F(\lambda_0-\varepsilon/2)-F(\lambda_0-3\varepsilon/2)
\big]\nonumber,\\
c_4&=&\frac{1}{4!\varepsilon^4}\big[
F(\lambda_0+2\varepsilon)-4~F(\lambda_0+\varepsilon)+6~F(\lambda_0)
-4~F(\lambda_0-\varepsilon)+F(\lambda_0-2\varepsilon)
\big]. \nonumber 
\end{eqnarray}
Comparing coefficients of $\lambda$-expansions in 
$S(\lambda~\eta , \lambda~\beta)$ and $1/G(\lambda \eta)$ 
(cf. Table \ref{tab2}), we confirm that eq.(\ref{LAMBDA-EXPAN}) 
approximately holds. 
This in turn means that the $\lambda$ dependences of $G(\lambda
\eta)$ and $S(\lambda~\eta , \lambda~\beta)$ in the l.h.s. of
eq.(\ref{SCALING}) effectively cancel out and, as the results, the
quasiscaling behavior emerges. The same is true for the exponential
source function.\\ 

In Tables \ref{tab3},~\ref{tab4} we present results of same examinations of
$S(\lambda~\eta_{+},\lambda~\beta)$ and $1/G(\lambda \eta_{+})$ but
for the production of $K^+$ meson instead (Table \ref{tab3} contains
results of the the Taylor series expansion and Table \ref{tab4} those 
for the finite
difference method). 
As seen in Tables \ref{tab3},~\ref{tab4}, they confirm that coefficients in
$1/G(\lambda \eta)$ agree with those of $S(\lambda~\eta ,
\lambda~\beta)$ also for the $K$ meson production case. Notice that
eq.(\ref{LAMBDA-EXPAN}) holds well for large $m_{\scriptscriptstyle
T}$-$m_{\pi(K)}$ for both $N^{\pi^{+}}$ and $N^{K^{+}}$ (this is
clearly seen in the first order $\varepsilon$ of $\lambda$=2 in 
Tables I~-~IV).\\

We examine dependence of the coefficients of
$S(\lambda\eta_{+},\lambda\beta)$ and  $1/G(\lambda \eta_{+})$ 
expansions on the size of $\varepsilon$ (both for the Gaussian and
for the exponential source functions), cf. Table \ref{tab5} 
and \ref{tab6}~($S(\lambda~\eta_{+},\lambda~\beta)$ calculated 
using the Gaussianand the exponential source functions are denoted by 
$S_g$ and $S_e$,respectively). Expansions are done around 
$\lambda_0$(=2) with fixed
values of $m_{\scriptscriptstyle T}$-$m_{\pi}$ or
$m_{\scriptscriptstyle T}$-$m_{K}$ ($=~200$ MeV in both cases) with
$\varepsilon = 1/2,~1/20$ and $1/200$, respectively. 
It can be observed that 
coefficients of the $\lambda$-expansion in $1/G(\lambda \eta_{+})$
agree reasonably well with those for
$S(\lambda\eta_{+},\lambda\beta)$ up to $\varepsilon =1/2$. This
means that quasiscaling behavior holds not only on the local scale
(i.e., for small $\varepsilon$) but also globally (up to $\varepsilon$
of the order of $1/2$). \\ 

\subsection{Numerical examination of the quasiscaling behavior}
To visualize the global structure of
the $\beta$-$Z_{\mbox{\scriptsize eff}}$ quasiscaling more clear
we examine the $\lambda$-dependence of the following
quantity:
\begin{eqnarray}
D^{\pi^+}(\lambda)
\equiv \frac{1}{\varepsilon}~
\Big[~N^{\pi^{+}}(
~(\lambda+\varepsilon/2)\eta,
~(\lambda+\varepsilon/2)\beta)-
N^{\pi^{+}}(
~(\lambda-\varepsilon/2)\eta,
~(\lambda-\varepsilon/2)\beta)~\Big]\label{DIFF},  
\end{eqnarray}
Figure 2 shows the $\lambda$ dependence of the
$D^{\pi^+}(\lambda;~m_{\scriptscriptstyle T}-m_{\pi})$ for both
the Gaussian $(a)$ and the exponential $(b)$ source functions used for
$N^{\pi^+}$. Figure 3 shows the same but for the $K^+$ meson 
production instead,
i.e., the $\lambda$ dependence of the
$D^{K^+}(\lambda;~m_{\scriptscriptstyle T}-m_{K})$.  
Except for the cases in which either $\lambda \ll 1$ or
$m_{\scriptscriptstyle T}$-$m_{\pi(K)}$ is small,
$D^{\pi^+(K^+)}(\lambda;~m_{\scriptscriptstyle T}-m_{\pi})$ is
essentially equal zero. In these figures we can readily observe
disappearance of the $\lambda$ dependence of the l.h.s. of
eq.(\ref{SCALING}) and the emergence of the quasiscaling behavior.
On the other hand we can also observe that for small values of
$m_{\scriptscriptstyle T}-m_{\pi(K)}$ the quasiscaling breaks down
because $D(\lambda;~m_{\scriptscriptstyle T}-m_{\pi}) \ne 0 $
there \cite{FOOT5}.\\ 

\section{How to estimate the size of the CCF}
It is tempting to use our results to estimate the size of CCF
as given by the effective number of stopped protons,
$Z_{\mbox{\scriptsize eff}}$. To this end we need some additional
inputs, which can be provided in two ways: Either by adopting 
empirical geometrical picture of the nucleus or 
by utilizing results obtained from the HBT effect.\\

First we use the empirical
formula for the radius of a nucleus (of an atomic number $A$),
\begin{eqnarray} 
\langle r_{\pi}^2 \rangle^{1/2} 
\approx r_0~A^{1/3}\label{EMPIRICAL},
\end{eqnarray}
where $r_0 = 1.1 \sim 1.2$ and, in our case, $A \approx (2.5 \sim 3.0)
\times Z_{\mbox{\scriptsize eff}}$, because of 
A=207 and Z=82. We use the root mean square
radius of the pionic source as $\langle r_{\pi}^2 \rangle^{1/2}\equiv
\{4\pi \int dr r^4 \rho(r)\}^{1/2}$ (which in terms of the source
size parameter $\beta$ used before is given as $\langle r_{\pi}^2
\rangle^{1/2}=\!\!\sqrt{3}\beta$ and $\langle r_{\pi}^2
\rangle^{1/2}=\!\!2\sqrt{3}\beta$ for the Gaussian and the exponential
sources, respectively.). 
Using $\langle r_{\pi}^2 \rangle^{1/2}$ instead of $\beta$, 
we can rewrite Fig.1. cf. Fig.4. 
Dotted line shows there the relation 
\begin{eqnarray}
Z_{\mbox{\scriptsize eff}}=\frac{1}{2.5r_0^3}\langle r_{\pi}^2 \rangle^{3/2},
\label{NORMAL-MATTER}
\end{eqnarray}
which provides us with $Z_{\mbox{\scriptsize eff}} = 45 \sim 60$ for
$\langle r_{\pi}^2 \rangle^{1/2} \approx 6$~fm (where both lines
cross each other).\\

In the second method we first apply the following 
standard formula with the Gaussian source function to the data of Pb+Pb
collisions at energy 158GeV/nucleon \cite{FRANZ}:
\begin{eqnarray}
N^{\pm\pm}/N^{BG}=c~(1+\lambda e^{-\beta^2Q^2}), 
\end{eqnarray}
(here $c$ is a normalization constant and $\lambda$ is a degree of
coherence parameter). As a result we obtain the following sets of
parameters for $\pi^+\pi^+$ and $\pi^-\pi^-$: \\($c, \lambda,
\beta$)=(1.00$\pm$0.01,~0.55$\pm0.03$,~6.02$\pm$0.35) with
$\chi^2$/NDF=13.0/17 and\\ ($c, \lambda,
\beta$)=(1.01$\pm$0.01,~0.47$\pm0.03$,~5.58$\pm$0.40)  
with $\chi^2$/NDF=7.4/17, respectively. Therefore, the pionic source
size of Pb+Pb collisions is estimated to be equal $12.7$~fm$\le\langle
r_{\pi}^2 \rangle^{1/2}\le15.6$~fm. Using this interval 
as the size of the CCF, we estimate
that $125 \le Z_{\mbox{\scriptsize eff}}\le 150$, which is much larger
than the that of CCF obtained from the empirical geometrical picture. 
It means that the CCF represents an expanding source, 
provided that the estimation by means of HBT effect is correct.
From our estimations, the baryon number density 
\begin{eqnarray}
~2.5\times150~\Big/~\frac{4\pi (~15.1~~\mbox{fm})^3}{3}&\!\! \le \!\!& 
n_{\mbox{\scriptsize B}}\le 
~2.5\times125~\Big/~\frac{4\pi (~12.5~~\mbox{fm})^3}{3}\nonumber \\
0.024 &\!\!\le \!\!& n_{\mbox{\scriptsize B}}\le 0.036 ~\quad[1/\mbox{fm}^3]
\end{eqnarray}
is obtained. \\

\section{Concluding remarks}

The mathematical structure for the $\beta$-$Z_{\mbox{\scriptsize
eff}}$ quasiscaling in eq.(\ref{SCALING}) found previously in
\cite{YIELD-RATIO} is examined in more detail. In particular,
through the Taylor series expansion (up to third order) we find that 
coefficients in the $\lambda$-expansion of $1/G(\lambda \eta)$  
are approximately same as those of $S(\lambda~\eta , \lambda~\beta)$. 
Using the finite difference method we can improve our examination 
by extending it to the fourth or fifth orders, with similar result.
This means that eq.(\ref{LAMBDA-EXPAN}) holds indeed (albeit only
approximately) and that $\lambda$-dependence of the l.h.s. of
eq.(\ref{SCALING}) is cancelled out in the product of $G(\lambda
\eta)$ and $S(\lambda~\eta , \lambda~\beta)$ leading in consequence
to the quasiscaling behavior observed. 
Finally we estimate the $Z_{\mbox{\scriptsize eff}}$ 
of the CCF in two possible ways.
From our estimations, the baryon number density 
$0.024 \le n_{\mbox{\scriptsize B}} \le 0.036 ~~[~1/\mbox{fm}^3~]$
is obtained. \\


\acknowledgements
One of the authors (T.O.) would like to
thank many people who supported him at the Department of Physics of
Shinshu University. The other author (G.W.) would like to thank the
Yamada Foundation for financial support and Department of Physics of
Shinshu University for hospitality extended to him during his visit
there. This work is partially supported by Japanese Grant-in-Aid for
Science  Research from the Ministry of Education, Science and Culture
(\#.~06640383) and (\#.~08304024). \\

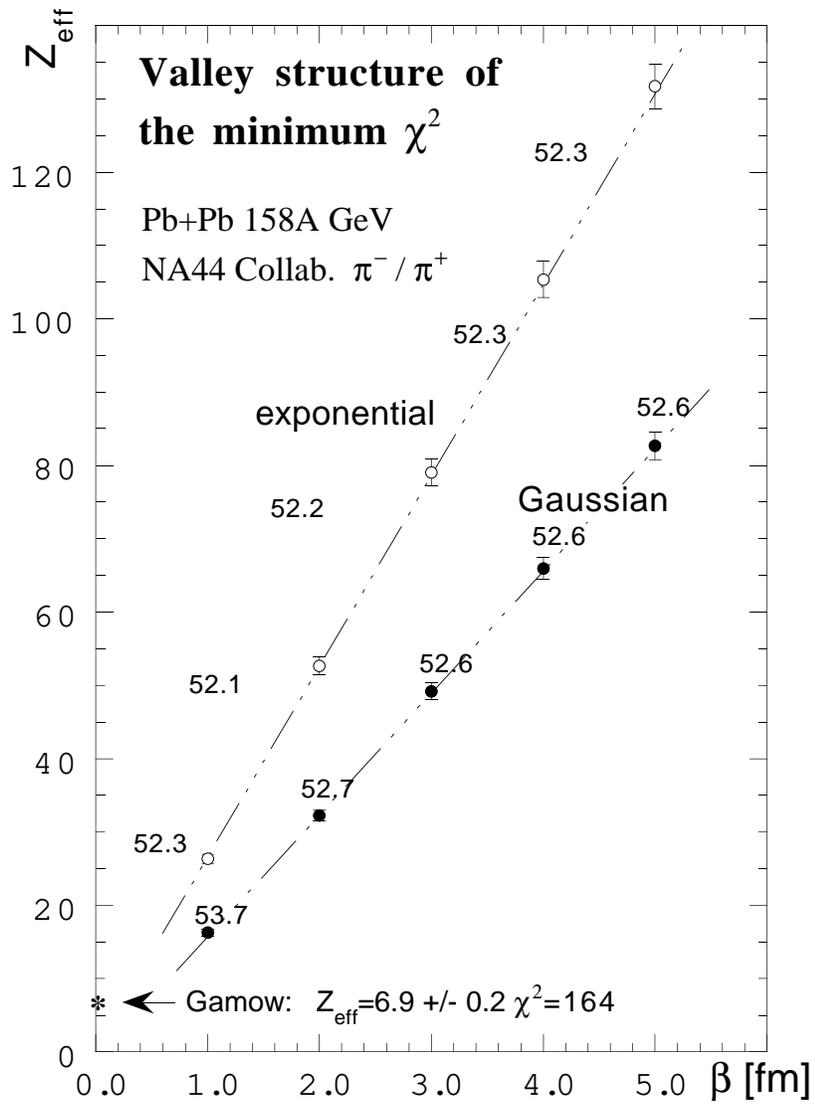
\begin{figure}
\caption{Valley structure of the minimum $\chi^2$ values in 
$Z_{\scriptscriptstyle \rm{eff}}$-$\beta$ parameter space 
for NA44 Collab.[2] 
data for the Gaussian and exponential pionic source functions, 
respectively.}
\label{fig1}
\end{figure}

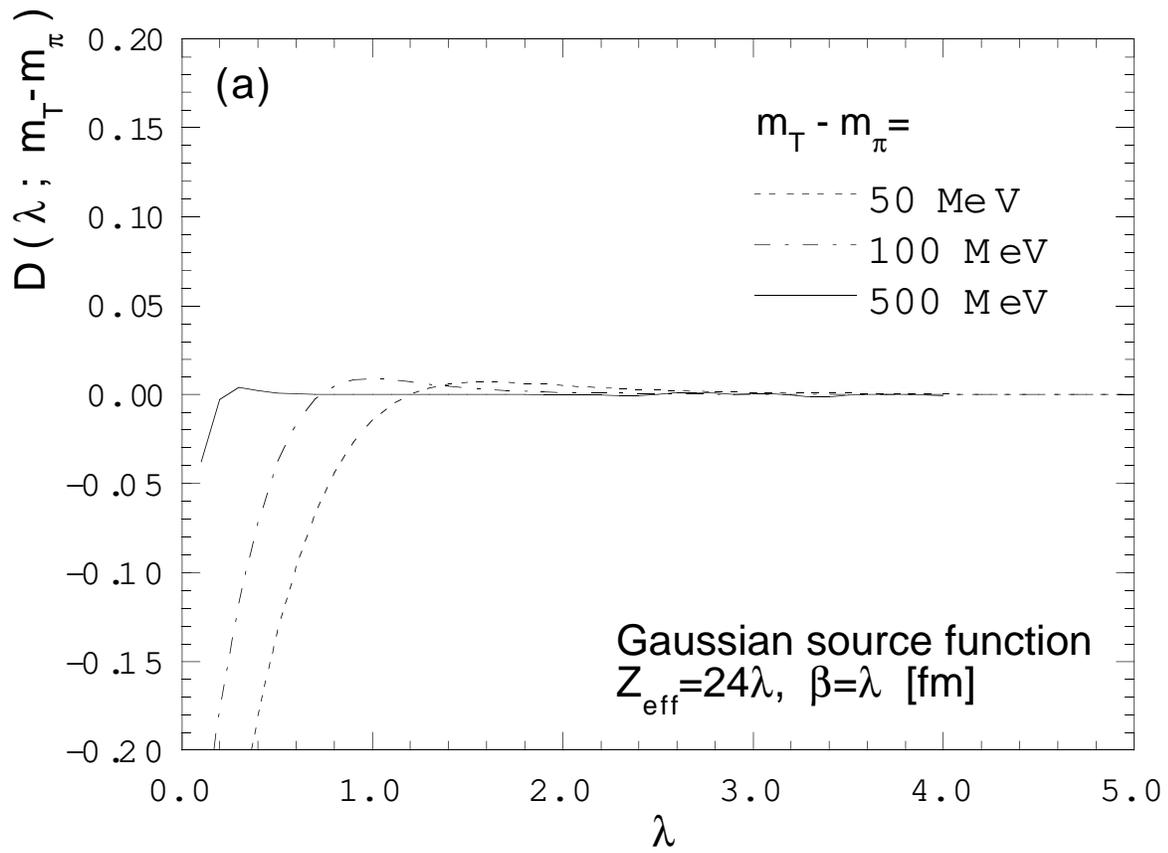
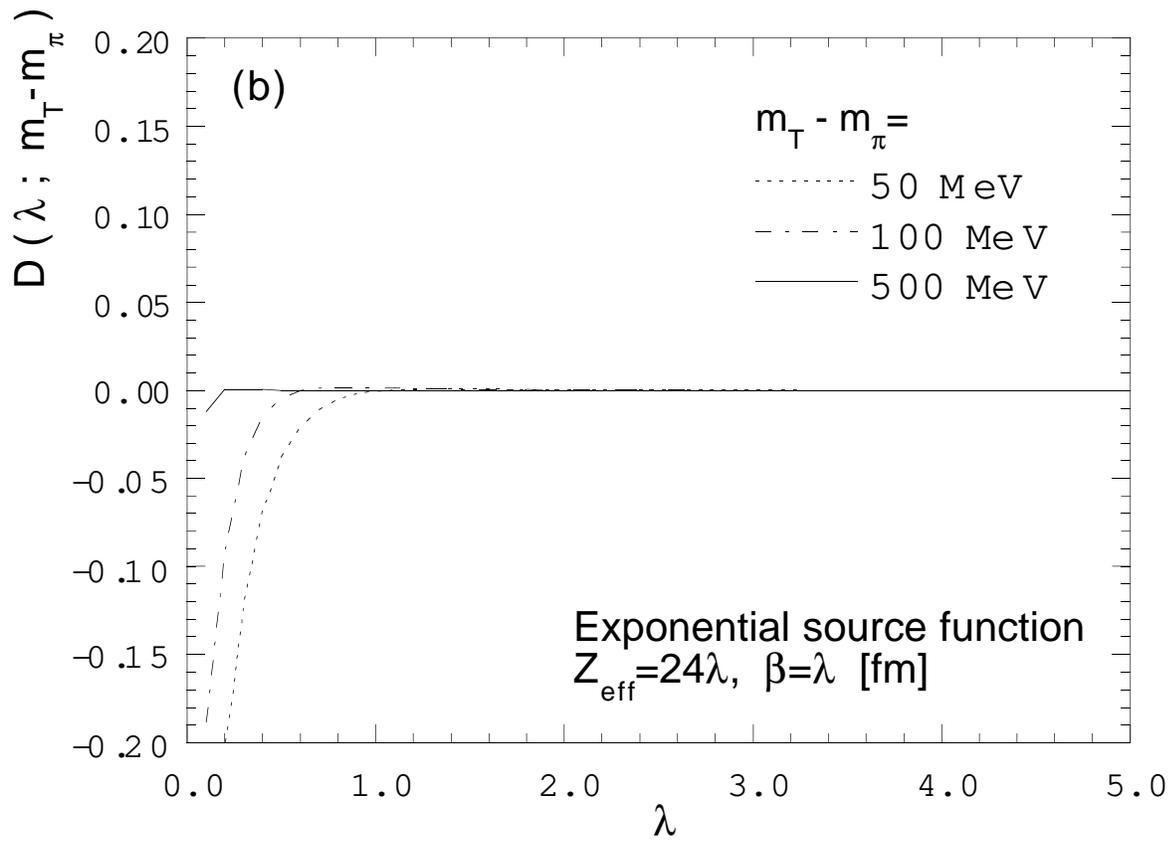
\begin{figure}
\caption{
Numerical examination of the 
$Z_{\scriptscriptstyle \rm{eff}}$-$\beta$ quasiscaling in $N^{\pi^{+}}
(m_{\scriptscriptstyle T}-m_{\pi}; 
~\lambda\!\times\!\eta,~\lambda\!\times\!\beta)$~for the Gaussian $(a)$
and exponential $(b)$ source functions.}
\label{fig2}
\end{figure}

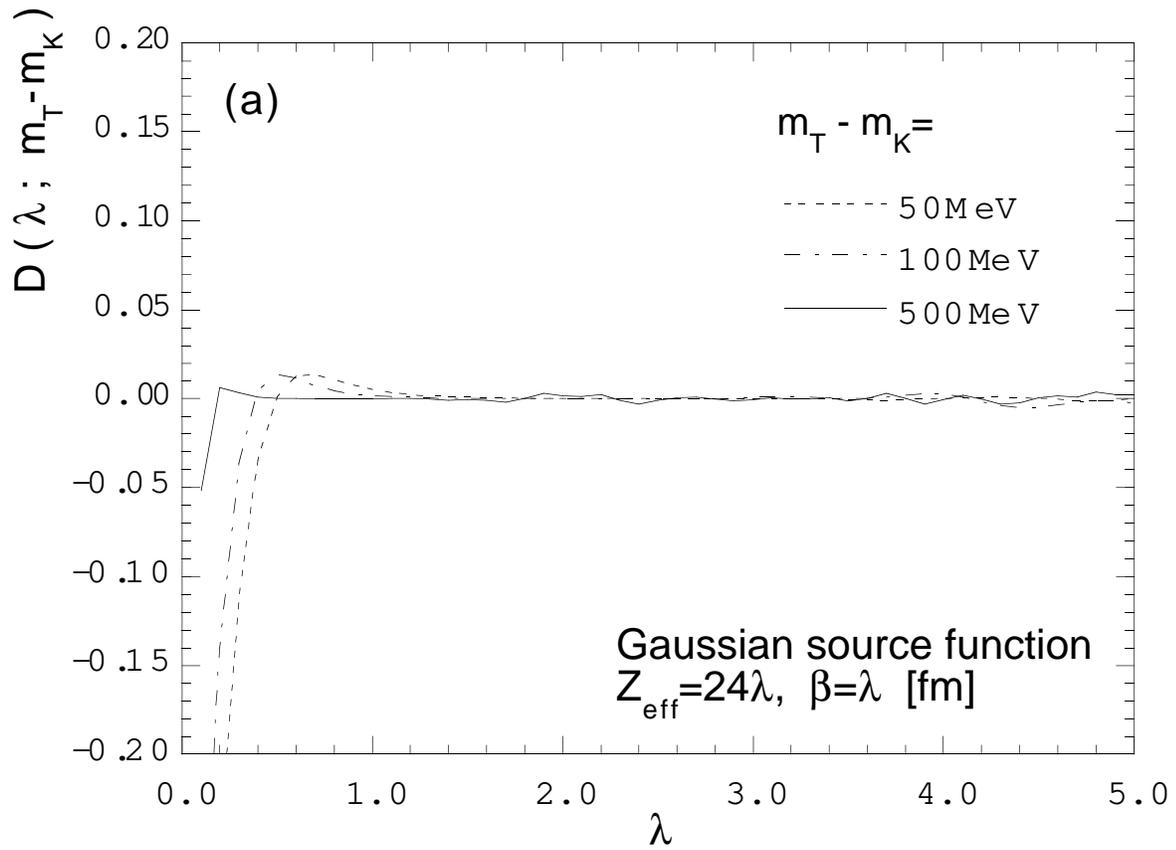
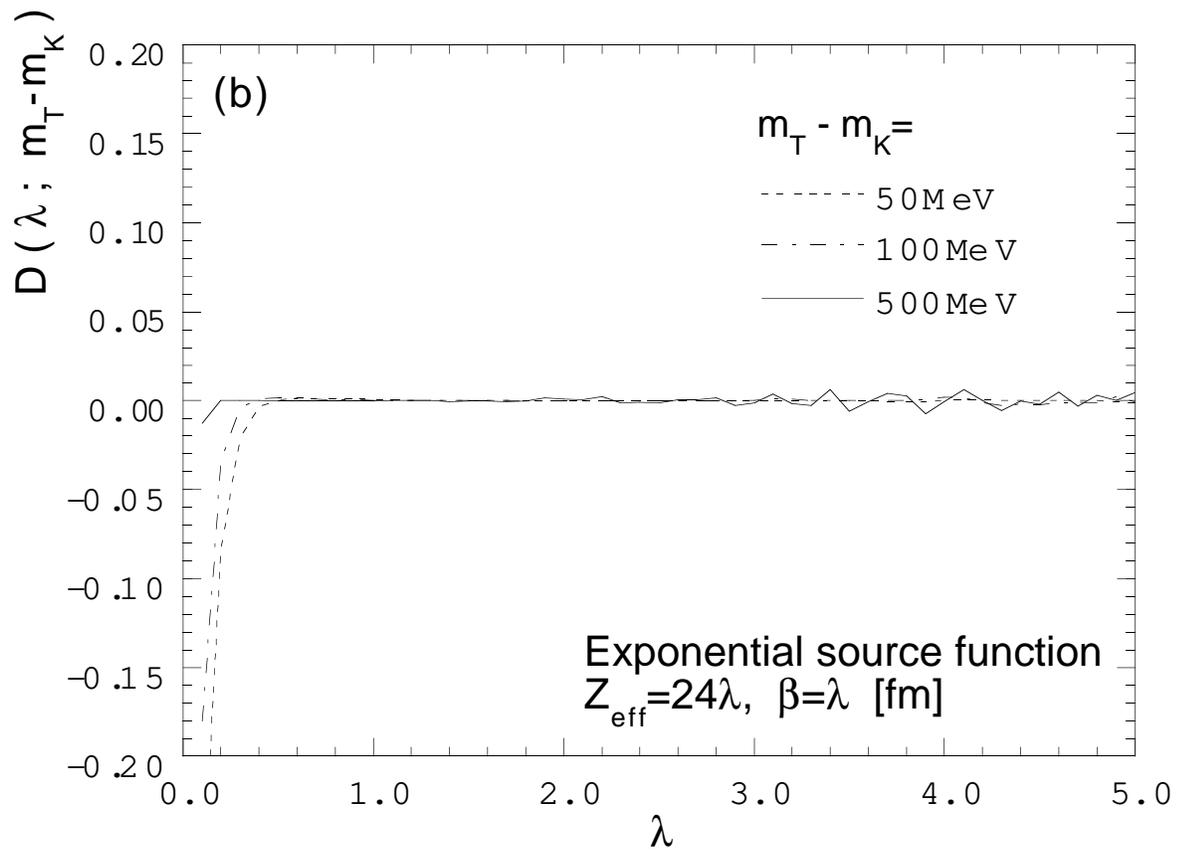
\begin{figure}
\caption{The same as in Fig. 2 but for the production of $K$ mesons.}
\label{fig3}
\end{figure}

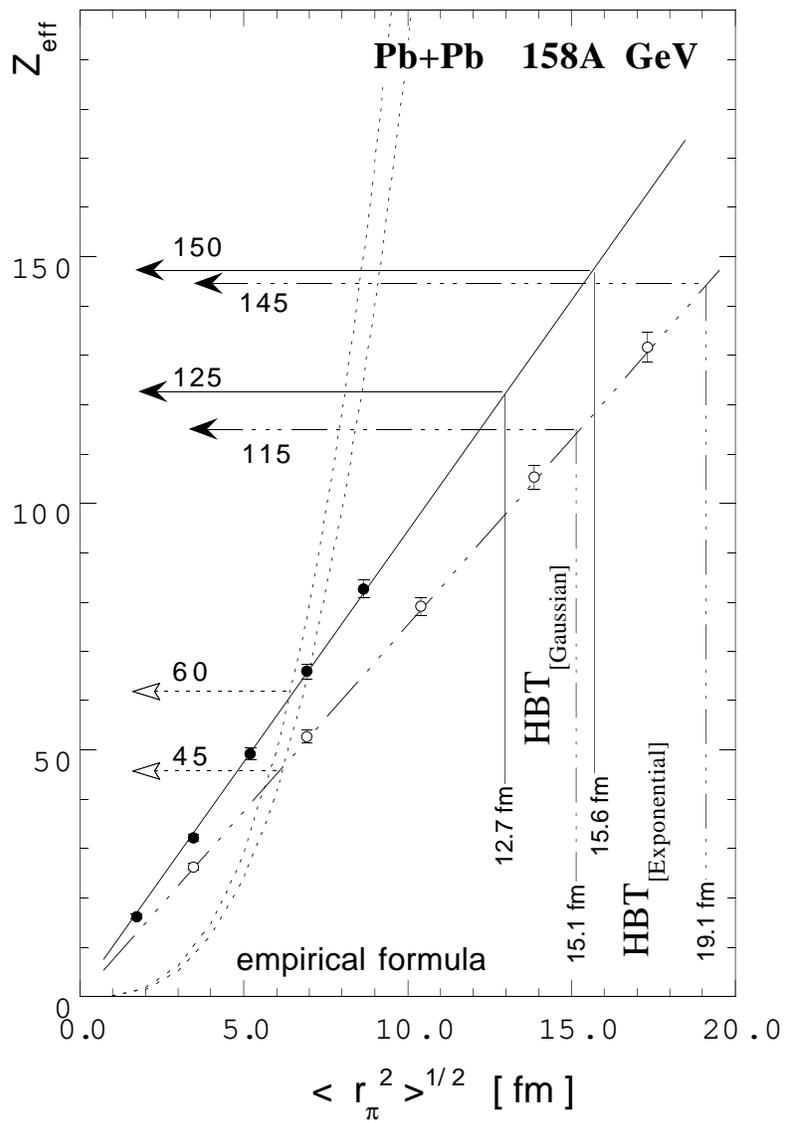
\begin{figure}
\caption{Two lines in Fig.1(for the Gaussian and the exponential 
source function) are rewritten by using of the r.m.s. radius 
$\langle r_{\pi}^2\rangle^{1/2}$. The dotted line is evaluate by
eq.(11). }
\label{fig4}
\end{figure}
\clearpage 

\begin{table}
\caption{
Comparison of the coefficients of the $\lambda$-expansion 
(the Taylor series) in $1/G(\lambda \eta)$ and in 
$S(\lambda~\eta , \lambda~\beta)$ around the $\lambda_0$ 
with fixed $m_{\scriptscriptstyle \rm{T}}$-$m_{\pi}$; 
$\delta\lambda=~\mid\lambda-\lambda_0 \mid \ll \varepsilon$. 
The Gaussian source function is used in
$S(\lambda~\eta , \lambda~\beta)$. 
In this Table we omit imaginary parts of coefficients in the
expansion of $S(\lambda~\eta , \lambda~\beta)$ emerging from the
limited computer ability. For example, in $\lambda$=1,
$m_{\scriptscriptstyle \rm{T}}$-$m_{\pi}$=50 MeV case, we obtain in
reality following result at out computer facility:
$S(\lambda \eta, \lambda \beta)=1.6531+i~1.1956\times 10^{-18}
+(0.8622+i~1.6171\times 10^{-17})~\epsilon
+(0.3853+i~1.0199\times 10^{-16})~\epsilon^2 +\cdots~.$}
\label{tab1}
\begin{tabular}{ll}
$N^{\pi^{+}}$& $m_{\mbox{\scriptsize T}}$-$m_{\pi}$\\
\hline
$\lambda_0 =1 $   & 50~MeV \\ 
1/G($\lambda \eta$)&$=1.6531+0.8902~\delta\lambda+0.2971~\delta\lambda^2+\cdots$ \\
$S(\lambda \eta, \lambda \beta)$&$ =1.6531+0.8622~\delta\lambda
+0.3853~\delta\lambda^2 +\cdots$ \hspace*{4.0cm}\\ 
\hline
$\lambda_0 =1 $   & 100~MeV \\ 
1/G($\lambda \eta$)&$=1.4129+0.5134~\delta\lambda+0.1180~\delta\lambda^2+\cdots$ \\
$S(\lambda \eta, \lambda \beta)$&$ =1.4126+0.5274~\delta\lambda
+0.1217~\delta\lambda^2 +\cdots$ \\ 
\hline
$\lambda_0 =1 $   & 200~MeV \\ 
1/G($\lambda \eta$)&$=1.2468+0.2845~\delta\lambda+0.4180~\delta\lambda^2+\cdots$ \\
$S(\lambda \eta, \lambda \beta)$&$ =1.2467+0.2873~\delta\lambda
+0.3674~\delta\lambda^2 +\cdots$ \\ 
\hline
$\lambda_0 =2 $   & 50~MeV \\ 
1/G($\lambda \eta$)&$=2.9283+1.7687~\delta\lambda+0.6241~\delta\lambda^2+\cdots$ \\
$S(\lambda \eta, \lambda \beta)$&$ =2.9272+1.7866~\delta\lambda
+0.6238~\delta\lambda^2 +\cdots$ \\ 
\end{tabular} 
\end{table} 

\begin{table}
\caption{
Comparison of the coefficients of the $\lambda$-expansion 
(the finite difference method, $\varepsilon = 1/200$) 
in $S(\lambda~\eta , \lambda~\beta)$ around the $\lambda_0$ 
with fixed $m_{\scriptscriptstyle \rm{T}}$-$m_{K}$. 
The Gaussian source function is used in 
$S(\lambda~\eta ,\lambda~\beta)$.}\label{tab2}
\begin{tabular}{ll}
$N^{\pi^{+}}$& $m_{\mbox{\scriptsize T}}$-$m_{\pi}$ \\
\hline
$\lambda_0 =2 $   & 50~MeV \\ 
1/G($\lambda \eta$)&$=2.9283+1.7687~\delta\lambda+0.6242~\delta\lambda^2 
+0.1572~\delta\lambda^3 +0.0309~\delta\lambda^4 +\cdots$ \\
$S(\lambda \eta, \lambda \beta)$&$=2.9283+1.7867~\delta\lambda
+0.6239~\delta\lambda^2 +0.1555~\delta\lambda^3 +0.0358~\delta\lambda^4 +\cdots$ \\ 
\hline
$\lambda_0 =2 $   & 100~MeV \\ 
1/G($\lambda \eta$)&$=2.0672+0.8216~\delta\lambda+0.1973~\delta\lambda^2 
+0.0342~\delta\lambda^3 +0.0047~\delta\lambda^4+\cdots$ \\
$S(\lambda \eta, \lambda \beta)$&$=2.0672+0.8245~\delta\lambda
+0.1958~\delta\lambda^2 +0.0356~\delta\lambda^3 +0.0037~\delta\lambda^4 +\cdots$ \\ 
\hline
$\lambda_0 =2 $   & 200~MeV \\ 
1/G($\lambda \eta$)&$=1.5780+0.3834~\delta\lambda+0.0581~\delta\lambda^2 
+0.0064~\delta\lambda^3 +0.0006~\delta\lambda^4+\cdots$ \\
$S(\lambda \eta, \lambda \beta)$&$=1.5780+0.3837~\delta\lambda
+0.0579~\delta\lambda^2 +0.0066~\delta\lambda^3 +0.0005~\delta\lambda^4 +\cdots$ \\ 
\end{tabular} 
\end{table} 

\begin{table}
\caption{The same as in Table I 
but for the production of $K$ mesons.}
\label{tab3}
\begin{tabular}{ll}
$N^{K^{+}}$& $m_{\mbox{\scriptsize T}}$-$m_{K}$\\
\hline
$\lambda_0 =1 $   & 50~MeV \\ 
1/G($\lambda \eta$)&$=2.2584+2.0291~\delta\lambda+1.0916~\delta\lambda^2+\cdots$ \\
$S(\lambda \eta, \lambda \beta)$&$ =2.2584+2.0459~\delta\lambda
+1.0777~\delta\lambda^2 +\cdots$ \hspace*{4.0cm}\\ 
\hline
$\lambda_0 =1 $   & 100~MeV \\ 
1/G($\lambda \eta$)&$=1.9074+1.3380~\delta\lambda+0.5723~\delta\lambda^2+\cdots$ \\
$S(\lambda \eta, \lambda \beta)$&$ =1.9074+1.2893~\delta\lambda
-1.9604~\delta\lambda^2 +\cdots$ \\ 
\end{tabular} 
\end{table} 

\begin{table}
\caption{The same as in Table II 
but for the production of $K$ mesons.}
\label{tab4}
\begin{tabular}{ll}
$N^{K^+}$& $m_{\mbox{\scriptsize T}}$-$m_{K}$ \\
\hline
$\lambda_0 =1 $   & 100~MeV \\ 
1/G($\lambda \eta$)&$=1.9074+1.3381~\delta\lambda+0.5723~\delta\lambda^2 
+0.1774~\delta\lambda^3 +0.0432~\delta\lambda^4 +\cdots$ \\
$S(\lambda \eta, \lambda \beta)$&$=1.9074+1.3417~\delta\lambda
+0.5669~\delta\lambda^2 +0.1866~\delta\lambda^3 +0.0298~\delta\lambda^4 +\cdots$ \\ 
\hline
$\lambda_0 =2 $   & 50~MeV \\ 
1/G($\lambda \eta$)&$=5.9708+6.2065~\delta\lambda +3.5872~\delta\lambda^2 
+1.4529~\delta\lambda^3 +0.4551~\delta\lambda^4 +\cdots$ \\
$S(\lambda \eta, \lambda \beta)$&$=5.9708+6.2079~\delta\lambda
+3.5864~\delta\lambda^2 +1.4531~\delta\lambda^3 +0.4541~\delta\lambda^4 +\cdots$ \\ 
\hline
$\lambda_0 =2 $   & 100~MeV \\ 
1/G($\lambda \eta$)&$=4.0487+3.2420~\delta\lambda+1.4789~\delta\lambda^2 
+0.4770~\delta\lambda^3 +0.1195~\delta\lambda^4 +\cdots$ \\
$S(\lambda \eta, \lambda \beta)$&$=4.0487+3.2420~\delta\lambda
+1.4786~\delta\lambda^2 +0.4777~\delta\lambda^3 +0.1189~\delta\lambda^4 +\cdots$ \\ 
\hline
$\lambda_0 =2 $   & 200~MeV \\ 
1/G($\lambda \eta$)&$=2.7398+1.5454~\delta\lambda+0.5123~\delta\lambda^2 
+0.1215~\delta\lambda^3 +0.0225~\delta\lambda^4 +\cdots$ \\
$S(\lambda \eta, \lambda \beta)$&$=2.7398 +1.5463~\delta\lambda
+0.5172~\delta\lambda^2 +0.1296~\delta\lambda^3 +0.0022~\delta\lambda^4 +\cdots$ \\ 
\end{tabular} 
\end{table} 

\begin{table}
\caption{Dependence of $\lambda$ expansion of $1/G(\lambda \eta)$,
$S_g(\lambda~\eta , \lambda~\beta)$  and $S_e(\lambda~\eta ,
\lambda~\beta)$ for $\pi^+$ production (in finite difference method)
on the size of $\varepsilon$ (for fixed values of $\lambda_0 = 2$ and
$m_{\scriptscriptstyle \rm{T}}-m_{\pi}$ = 200 MeV.)}
\label{tab5}
\begin{tabular}{ll}
$N^{\pi^{+}}$& $m_{\mbox{\scriptsize T}}$-$m_{\pi}$
;~(~$\beta$=2~fm, $Z_{\mbox{\scriptsize eff}}$=48~) \\
\hline 
$\varepsilon=1/2$   & 200~MeV\\ 
1/G($\lambda \eta$)&$=1.5780+0.3838~\delta\lambda+0.0582~\delta\lambda^2 
+0.0065~\delta\lambda^3 +0.0006~\delta\lambda^4+\cdots$ \\
$S_g(\lambda \eta, \lambda \beta)$&$=1.5780+0.3841~\delta\lambda
+0.0580~\delta\lambda^2 +0.0067~\delta\lambda^3+0.0004~\delta\lambda^4+ \cdots$ \\ 
$S_e(\lambda \eta, \lambda \beta)$&$=1.5780+0.3839~\delta\lambda
+0.0581~\delta\lambda^2 +0.0065~\delta\lambda^3+0.0005~\delta\lambda^4+ \cdots$ \\ 
\hline
$\varepsilon=1/20$   & 200~MeV\\ 
1/G($\lambda \eta$)&$=1.5780+0.3834~\delta\lambda+0.0581~\delta\lambda^2 
+0.0064~\delta\lambda^3 +0.0006~\delta\lambda^4+\cdots$ \\
$S_g(\lambda \eta, \lambda \beta)$&$=1.5780+0.3837~\delta\lambda
+0.0579~\delta\lambda^2 +0.0066~\delta\lambda^3+0.0005~\delta\lambda^4+ \cdots$ \\ 
$S_e(\lambda \eta, \lambda \beta)$&$=1.5780+0.3835~\delta\lambda
+0.0580~\delta\lambda^2 +0.0068~\delta\lambda^3+0.0039~\delta\lambda^4+ \cdots$ \\ 
\hline
$\varepsilon=1/200$   & 200~MeV\\
1/G($\lambda \eta$)&$=1.5780+0.3834~\delta\lambda+0.0581~\delta\lambda^2 
+0.0064~\delta\lambda^3 +0.0006~\delta\lambda^4+\cdots$ \\
$S_g(\lambda \eta, \lambda \beta)$&$=1.5780+0.3837~\delta\lambda
+0.0579~\delta\lambda^2 +0.0066~\delta\lambda^3 +0.0005~\delta\lambda^4 +\cdots$ \\ 
$S_e(\lambda \eta, \lambda \beta)$&$=1.5780+0.3835~\delta\lambda
+0.0579~\delta\lambda^2 +0.0069~\delta\lambda^3 +0.0047~\delta\lambda^4 +\cdots$ \\ 
\end{tabular} 
\end{table} 

\begin{table}
\caption{The same as in Table V but for the production of $K$ mesons.}
\label{tab6}
\begin{tabular}{ll}
$N^{K^{+}}$& $m_{\mbox{\scriptsize T}}$-$m_{K}$
;~(~$\beta$=2~fm, $Z_{\mbox{\scriptsize eff}}$=48~) \\
\hline
$\varepsilon=1/2$   & 200~MeV\\ 
1/G($\lambda \eta$)&$=2.7398+1.5530~\delta\lambda+0.5180~\delta\lambda^2 
+0.1236~\delta\lambda^3 +0.0231~\delta\lambda^4+\cdots$ \\
$S_g(\lambda \eta, \lambda \beta)$&$=2.7398+1.5541~\delta\lambda
+0.5189~\delta\lambda^2 +0.1219~\delta\lambda^3+0.0200~\delta\lambda^4+ \cdots$ \\ 
$S_e(\lambda \eta, \lambda \beta)$&$=2.7398+1.5535~\delta\lambda
+0.5181~\delta\lambda^2 +0.1225~\delta\lambda^3+0.0222~\delta\lambda^4+ \cdots$ \\ 
\hline
$\varepsilon=1/20$   & 200~MeV\\ 
1/G($\lambda \eta$)&$=2.7398+1.5455~\delta\lambda+0.5124~\delta\lambda^2 
+0.1215~\delta\lambda^3 +0.0225~\delta\lambda^4+\cdots$ \\
$S_g(\lambda \eta, \lambda \beta)$&$=2.7398+1.5464~\delta\lambda
+0.5172~\delta\lambda^2 +0.1292~\delta\lambda^3+0.0017~\delta\lambda^4+ \cdots$ \\ 
$S_e(\lambda \eta, \lambda \beta)$&$=2.7398+1.5453~\delta\lambda
+0.5140~\delta\lambda^2 +0.1504~\delta\lambda^3+0.0160~\delta\lambda^4+ \cdots$ \\ 
\hline
$\varepsilon=1/200$   & 200~MeV\\
1/G($\lambda \eta$)&$=2.7398+1.5454~\delta\lambda+0.5123~\delta\lambda^2 
+0.1215~\delta\lambda^3 +0.0225~\delta\lambda^4 +\cdots$ \\
$S_g(\lambda \eta, \lambda \beta)$&$=2.7398 +1.5463~\delta\lambda
+0.5172~\delta\lambda^2 +0.1296~\delta\lambda^3 +0.0022~\delta\lambda^4 +\cdots$ \\ 
$S_e(\lambda \eta, \lambda \beta)$&$=2.7398 +1.5453~\delta\lambda
+0.5139~\delta\lambda^2 +0.1541~\delta\lambda^3 +0.0130~\delta\lambda^4 +\cdots$ \\ 
\end{tabular} 
\end{table} 

\begin{thebibliography}{99}
\bibitem{FOOT1} It should be noted that in quantum mechanics
$N^{\pi^{\pm}}$ can be regarded as the expectation value of pion
production source function $\rho(r)$ for  the Coulomb wave function
$\psi_r^{\pm}({\bf p}_r,{\bf r})$:
\begin{eqnarray*}
N^{\pi^{\pm}}=\int dr [\psi_r^{\pm}({\bf p}_r,{\bf r})]^{\ast}~ 
\rho(r)~\psi_r^{\pm}({\bf p}_r,{\bf r}). 
\end{eqnarray*}
\bibitem{FOOT2} In what follows the results are presented only for
positively charged particles but all conclusions that follow are
equally valid for negatively charged ones.
\bibitem{FOOT4} 
This method needs a lot of CPU time and a vast computer
memory, therefore the respective Taylor series expansion parameters are
calculated here up the third order only.
\bibitem{FOOT5} Notice that in the case of $\lambda \ge 1.5 \sim
2.5$, it is difficult to evaluate $N^{\pi^{+}(K^+)}$ using series
expansion as in eq.(\ref{WAVE-INTEGRAL}). In this case we use the
seamless fitting method introduced in \cite{SEAMLESS}. 
For calculations of $N^{\pi^{\pm}}$ (eq.(\ref{SCALING})) for large 
$\lambda$, we also make use of the Whittaker function from the
CERN library, in order to calculate the hypergeometric 
function $F$ present
in the Coulomb wave function, cf. eq.(\ref{CONFHYP}).
\end{thebibliography}
\end{document}